# Understanding the influence of Individual's Self-efficacy for Information Systems Security Innovation Adoption: A Systematic Literature Review

Mumtaz Abdul Hameed

Technovation Consulting and Training Private, Limited 33, Chandhani Magu
Male'. Maldives
Email: mumtazabdulhameed@gmail.com

Nalin Asanka Gamagedara Arachchilage

School of Engineering and Information Technology

University of New South Wales, The Australian Defence Force Academy

Australia

Email: nalin.asanka@adfa.edu.au

## Abstract

Information Systems security cannot be fully apprehended if the user lacks the required knowledge and skills to effectively apply the safeguard measures. Knowledge and skills enhance one's self-efficacy. Individual self-efficacy is an important element in ensuring Information Systems safeguard effectiveness. In this research, we explore the role of individual's self-efficacy for Information Systems security adoption. The study uses the method of Systematic Literature Review using 42 extant studies to evaluate individual self- efficacy for Information Systems security innovation adoption. The systematic review findings reveal the appropriateness of the existing empirical investigations on the individual self-efficacy for Information Systems security adoption. Furthermore, the review results confirmed the significance of the relationship between individual self-efficacy and Information Systems security adoption. In addition, the study validates the past administration of the research on this subject in terms of sample size, sample subject and theoretical grounds.

## 1. Introduction

Information Systems (IS) assets (information and computer resources) are at risk from a variety of threats, including virus, worms, Trojans, spyware, scare-ware, crime-ware, key-loggers, botnet, DDoS, browser- hijackers, pharming, phishing etc. [8]. Such attacks commonly referred to as 'IS security threats' mainly intended to improperly disclose, modify or delete sensitive information and maliciously destruct and destroy computer resources [23]. New prospect the internet has presented to the users have in fact, offered criminals and individuals with a vicious mind-set to misuse IS assets aimlessly.

To thwart IS security threats and safeguard organisational IS assets in general, a combination of measures is taken such as the installation of anti-virus, anti-spyware and anti-phishing software, setting up firewalls, maintaining and restricting access controls, using intrusion detection and prevention systems and by putting in encryption and content filtering software [33, 38, 49]. These measures offer a technological or technical solution to the problem, but by no means reasonable to efficiently safeguard IS security threats completely [3, 30, 49, 50, 56, 73, 74]. So as to survive with increased threats and to effectively protect IS assets, non-technical solutions such IS security policies have likewise been employed [53]. Research has established the view that organisations and individuals who opt for technical as well as non-technical measures to protect their IS assets are more likely to attain success in safeguarding IS resources [47, 56, 65]. In IS literature an innovation is referred as an idea, a product, a process or a technology that is new to an individual or organisation [25, 27]. Hence, technical and non- technical IS security measures may collectively be referred as IS security innovations.

Although both technical and non-technical IS security measures are important, several research had pinpointed behaviour of individual user within an organization as one element of ambiguity in securing IS assets [5, 16, 56, 65]. With all the technical and non- technical IS security measures at one's disposal, efficient use cannot be realized if the end user lacks the required knowledge and skills to adequately apply the measures. If the end-users of organisational IS does not understand the importance of IS security practices and are not eager to accept the policies, then those safeguards measures become ineffective [30]. Given that the security attacks are increasingly widespread and more organized than ever, it is important to gauge the knowledge of users to detect and prevent such attacks.

When an individual possesses the necessary knowledge about the effectiveness of a safeguard measure in providing protection from IS threats, that individual is more likely to adopt preventive behaviour or action [38, 51, 68]. Chan et al. [10] stated that acquisition of knowledge related to an IS countermeasure builds one's self-confidence in dealing with threats. According to IS literature, computer self-efficacy is the term that relates individual's self- confidence and ability to successfully use a computer or IS to accomplish a specific task [4, 13]. Computer self- efficacy have also been cited as essential in determining one's intention to engage in current or future use of an IS.

Prior research on IS indicates a significant positive relationship between individual's IS self-efficacy and the usage of ISs [60]. Also, individuals IS self-efficacy has found to be a significant determinant for IS security adoption [30, 53, 66]. Eastin and LaRose [20] state that self-efficacy overcomes the fear many novice users experience in an event of threat and enhances the ability to cope with any attack. Arachchilage and Love [4], identified self-efficacy as an important determinant of the IS security threat avoidance behaviour and a key element in ensuring safeguard effectiveness.

This research attempts to examine the role of an individual's self-efficacy in IS security innovation adoption. To this end, the study reviewed past literature on IS security to establish the relationship between self- efficacy and IS security adoption. The research makes three main contributions to theory and practice. First, using a review of IS security literature, the research verifies the significance of examining the effect of individual self-efficacy on IS security adoption. Secondly, the analysis carried out established the existing savvy of the role of individual self-efficacy for IS security innovation adoption. Finally, the study approves the significance of individual self-efficacy for IS security innovation adoption.

The remainder of this paper is organised as follows. The 'Theoretical Background' section illustrates the basics of self-efficacy relating to IS security. In the subsequent section 'Research Questions', we presented 4 research questions for the study. The 'Research Methodology' section, briefly discusses the method employed to examine the influence of the relationship between self-efficacy and IS security innovation adoption. In Section 5, we presented the result obtained from the data analysis. Finally, in Section 6, we discussed the finding of the study results, in addition, conclusion was also presented in Section 6.

## 2. Theoretical Background

The focus of IS security is to protect and safeguard organization's IS assets from vulnerabilities [1]. The main challenge for organization's IS security is to protect unauthorized access of information sources [21] and to defend computer resources against malicious attacks. As a result, organizations allocate a substantial amount of resources to safeguard their IS assets from IS security threats [23].

Various solutions have been developed in response to IS security and these solutions targets both technical and non-technical problem areas [3, 5]. With all the IS security measures at one's disposal, the efficient use cannot be realised if the end user lacks the required knowledge and skills to adequately apply the measures. Banu and Banu [8] indicated that IS security attacks over the internet are successful because of many inexperienced and unsophisticated users. Additionally, social engineering attacks are now much more concealed as such naive users are more inclined to incautiously divulge passwords and other sensitive and classified information. Lack of awareness of the users regarding the maliciousness of crimes over the internet in effect has opened a fertile ground for cyber-criminals to conduct IS security attacks. Even in the present-day, a number of users are ignorant that their personal information is actively being targeted by cyber- criminals. Given that the security attacks are increasingly widespread and more organized than ever, it is important to develop the knowledge of users to detect and prevent such attacks.

According to Rogers [51], when individuals possess the requisite knowledge about the effectiveness of mechanisms that provide protection from a threat, they are more likely to adopt that measure. In other words, a person who is knowledgeable about IS security is more likely to assess IS security risks and accordingly employs security innovations effectively to address those risks [41]. Individual's knowledge has a co- relation to one's self-efficacy to perform a behaviour [3].

Bandura [7] defined self-efficacy as the judgment of one's ability to organize and execute given types of performance. Hence, in the context of this research, self-efficacy is referred as a belief in one's ability to thwart IS security threats and one's capability to safeguard IS assets from IS security attacks. Tamjidyamcholo et al. [59] noted that a high level of self-efficacy in a person will make them much more self-assured about their abilities and strengthens their motivation. Hence, when users are knowledgeable about IS security threats, they have more self-confidence to take relevant actions to thwart attack by adopting preventive behaviour.

Researchers often utilised Bandura [7]'s theory of self- efficacy to measure individual's self-confidence. The fundamental of this theory is in understanding the relationship between one's belief and one's willingness to engage in behaviours necessary to successfully accomplish a task. The theory also explains the process an individual experience as he or she encounters a new challenge together with the judgments, evaluations, and appraisals made based on the knowledge learnt [6].

## 3. Research Questions

This paper considered the existing IS security literature to determine the importance of individual self-efficacy for IS security innovation adoption. The analysis focused specifically on investigating, the following research questions:

RQ1: What are the demographics of the extant studies of individual self-efficacy on IS security innovation adoption including the year of study, sample groups, sample size, countries?

RQ2: What are the theoretical foundation used in the existing studies of individual self-efficacy on IS security innovation adoption?

RQ3: Is there a difference in investigating individual self-efficacy for different types of security innovations?

RQ4: What are the results of the studies that examine the relationship between individual self-efficacy and IS security innovation?

## 4. Research Methodology

A finding of an individual study is not sufficient to generalise on a particular issue, while to reach an overall outcome, findings of a number of independent studies on a subject can be combined [24]. A technique known as a Systematic Literature Review (SLR) may be used to identify, analyse and interpret all available evidence related to a specific research question [27]. To meet our research objectives and to address the research questions, we carried out a SLR to study the role of self- efficacy for IS security innovation adoption. SLR improves the likelihood of generating a clearer, more objective answer to the research questions. As SLRs considers study design (sampling strategy and data collection methods), data and analytical methods used, the reviews are effective at gauging the robustness of evidence. The use of SLR procedure enabled the study to obtain an overall conclusion regarding the relationships between individual self-efficacy and IS security adoption.

To ensure a thorough coverage of academic articles related to IS security adoption, we conducted an extensive literature search of IS Journals using Google Scholar and multiple large-scale and reputable digital libraries and databases including Web of Science, IEEE Xplore, Science Direct (Elsevier), ACM Digital Library, Wiley Online Library, ProQuest, EBSCO, Springer LINK and Emerald Management Xtra. These sources contain ample high-quality journal articles and conference papers. The search focused only on peer- reviewed journal and conference articles.

To determine which of the articles were really relevant to the research objectives the study established, an inclusion and exclusion conditions. The study selection criteria for the SLR were: (C1) it should be an empirical study on IS security adoption, (C2) the study should examine individual self-efficacy as a dependent variable, and finally, (C3) the study examines the relationship between individual self-efficacy and IS security innovation adoption.

The initial search yielded 544 citations by following inclusion and exclusion criterion C1. To accomplish the inclusion and exclusion criterion C2, the abstracts of all 544 were manually scanned to identify if the articles examine individual self-efficacy. Number of articles identified as potentially relevant were 112. By applying inclusion and exclusion criterion C3 for these 112 articles, 39 articles with 42 studies were found eligible for the SLR. The 42 studies that meet all 3 criteria examined the effect of individual self-efficacy for the adoption of IS security innovations.

## 5. Results

We conducted a statistical analysis using frequencies and percentages to combine and summarize the variables collected.

### 5.1. Distribution of Studies by Year

Table 1 shows the literature distribution by publication year of the studies. Data from the SLR shows that self- efficacy has been considered in the IS security innovation adoption literature since 2004.

| Year | No. of Studies |
|------|----------------|
| 2004 | 1 |
| 2005 | 1 |
| 2006 | 0 |
| 2007 | 3 |
| 2008 | 2 |
| 2009 | 8 |
| 2010 | 6 |
| 2011 | 2 |
| 2012 | 6 |
| 2013 | 3 |
| 2014 | 3 |
| 2015 | 0 |
| 2016 | 6 |

**Table 1: Literature distribution by publication year.**

The academic discussion of individual self-efficacy on IS security adoption has mostly taken place during the last 12 to 14 years. Table 1 shows that the number of articles over time has increased and during this period, the topic has increasingly attracted among the scholarly researchers. The distribution of studies by publication year suggests that examining individual self-efficacy for IS security innovation adoption is an increasingly emerging discourse. Also, SLR confirms that individual self-efficacy for IS security innovation adoption is still an active IS tract, as there were 6 articles published in the year 2016.

*5.2. Distribution of Sample Groups in the studies*
The result of this analysis provided some clarification to RQ1.

| Subject Groups | No of Studies |
|---|---|
| Individual | 18 |
| Organisation | 2 |
| Student | 18 |
| Mixed | 2 |
| None | 2 |

**Table 2: Distribution of sample groups used in the studies.**

Table 2 illustrates the number of studies that employ different sample groups in the studies to examine individual self-efficacy for IS security innovation adoption. Results suggest that the majority of studies conducted their studies by engaging individuals by adopting convenience sampling or by using student subjects. The analysis also helped explain RQ1.

*5.3. Distribution of Sample size in the studies*

SLR analysed sample size of the reviewed studies to further elucidate RQ1. Among the 42 studies considered in the SLR, 40 studies utilised survey methodology. In this 40 studies, a total of 13841 participants was included, with an average sample size of 346. Table 3 showed that the study employing smallest and largest sample were 77 and 988 participants, respectively. Approximately, 67% (two third) of the studies use more than 200 participants in their assessment.

| Description | No. of. |
|---|---|
| Studies with sample | 40 |
| Smallest sample size | 77 |
| Largest sample size | 988 |
| Sample Size 0 - 100 | 1 |
| Sample Size 101 - 200 | 12 |
| Sample Size 201 - 300 | 9 |
| Sample Size 301 - 400 | 3 |
| Sample Size 401 - 500 | 5 |
| Sample Size 501 - 600 | 4 |
| Sample Size 601 - 700 | 2 |
| Sample Size 701 - 800 | 1 |
| Sample Size 801 - 900 | 0 |
| Sample Size 901 - | 3 |

**Table 3: Distribution of sample size of the studies**.

*5.4. Distribution by countries*

As a final appraisal to RQ1, we analysed the moderating effect of the country of study. Table 4 visually indicates that almost half of the studies were produced in the USA. The studies covered Asia, Europe and North America with a representation of 8 different countries.

| Country | No. of Studies |
|---|---|
| Canada | 3 |
| China | 2 |
| Finland | 4 |
| Malaysia | 3 |
| Singapore | 2 |
| South Korea | 2 |
| Taiwan | 3 |

**Table 4: Distribution of country of the studies**

## 1.1. Theories Used in the Reviewed Studies

In response to RQ2, we analysed the theoretical foundation for each reviewed literature. To examine the relationship between self-efficacy and IS security innovation adoption, reviewed studies used a number of different theories. Table 5 shows the different theoretical model exploited in the reviewed studies.

PMT is the most widely used theory to determine the relationship between self-efficacy and IS security adoption. More than half of the reviewed studies used PMT or PMT integrated with other theories. Reviewed literature suggests that apart from PMT, the Theory of Planned Behaviour (TPB), Theory of Reasoned Action (TRA) and Social Cognitive Theory (SCT) are among the most widely used theories in examining self-efficacy on IS security innovation adoption.

## 1.2. Types of Innovation

According to the classification of Zmud [71] we defined the type of innovation as process and product. For this study, process innovation involves establishing a new system, method or policies that changes the IS security operational processes, whereas product innovation are new products introduced to enhance IS security. Different factors determine the adoption of process and product innovation and the extent to which these factors impact on the adoption process [61].

| Theories | No. of Studies |
|---|---|
| Protection Motivation Theory (PMT) | 22 |
| Theory of Planned Behaviour (TPB) | 6 |
| Theory of Reasoned Action (TRA) | 5 |
| Social Cognitive Theory (SCT) | 5 |
| Deterrence Theory (DT) | 4 |
| Technology Acceptance Model (TAM) | 3 |
| Technology Threat Avoidance Theory (TTAT) | 2 |
| Cognitive Evaluation Theory (CET) | 1 |
| Coping Theory (CT) | 1 |
| Decomposed Theory of Planned Behaviour (DTPB) | 1 |
| Extrinsic Motivational Model (EMM) | 1 |
| Health Belief Model (HBM) | 1 |
| Instrinsic Motivation Model (IMM) | 1 |
| Rational Choice Theory (RCT) | 1 |
| Social Bond Theory (SBT) | 1 |

We differentiate the reviewed studies into two sets of process and product innovation and examine some demographics including sample size, sample groups for each group of the studies. Also, we examine if there is any difference in the application of theories for the studies that examine process and product innovations. Table 6 highlights the difference in study practices for process and product security innovations. The result of this analysis would address to RQ3.

Also, it is evident from the results that most of IS security process innovation studies utilises individuals as a subject, whereas, most of IS security product innovation studies employs student participants.

**Table 5: Different theories used in the studies.**

| Description | Process | Product |
|---|---:|---:|
| No of Studies | 24 | 18 |
| Total sample size | 8954 | 4887 |
| Sample Group | | |
|     Individual | 13 | 5 |
|     Organisation | 2 | 0 |
|     Student | 8 | 10 |
|     Mixed | 0 | 2 |
|     None | 1 | 1 |
| Theories used | | |
|     Protection Motivation Theory (PMT) | 12 | 10 |
|     Theory of Planned Behaviour (TPB) | 4 | 2 |
|     Theory of Reasoned Action (TRA) | 5 | 0 |
|     Social Cognitive Theory (SCT) | 5 | 0 |
|     Deterrence Theory (DT) | 4 | 0 |
|     Technology Acceptance Model (TAM) | 2 | 1 |

**Table 6: Distribution of studies using different security innovations.**

As for the theories used for two groups of studies, process innovation studies tend to combine PMT with the theoretical basis of either TRA, SCT or Deterrence Theory (DT) compared to studies examining product innovations.

*5.7. Significance*

The relationship between independent and dependent variables is usually evaluated in term of 'test of significance', highlighting their relationship [25, 26]. 'Test of significance' and various other 'effect sizes' such as correlation co-efficient provided by quantitative studies can be aggregated to find an overall outcome [27]. Effect size when considered in terms of significance is frequently referred as weak, moderate or strong significance [24]. Hunter et al. [32] and Hameed and Counsell [25], however, suggested that aggregation of 'test of significance' results from different studies could produce a misleading outcome. This is because, there is no rule for determining the value of the correlation that interprets as weak, moderate or strong significance.

For the study, we extracted from the reviewed studies the correlation co-efficient values of the relationship between self-efficacy and IS security innovation adoption. We interpreted the correlation co-efficient values under a single classification to obtain the test of significance for our assessment. We adopted the correlation value referred by Hameed and Counsell [24] and Hameed and Counsell [26], which categorises: a correlation value between 0 and ±0.09 as insignificant, ±0.10 and ±0.29 as weak significance, ±0.30 and ±0.49 as moderate significance, ± 0.5 and ± 0.69 as strong significance, ±0.70 and ±0.89 as the very strong significance and ±0.9 and ±1.0 near perfect. Based on the above classification we coded the correlation co- efficient of individual studies and aggregated resulting tests of significance to obtain the overall assessment of the relationship between self-efficacy and IS security innovation adoption.

Among the 42 studies considered in the SLR, 35 studies provided correlation co-efficient for the relationship between individual self-efficacy and IS security innovation adoption. Table 7 summarizes the results of an aggregated test of significance for the relationship between self-efficacy and the adoption of IS security innovation.

| Significance | No. of Studies |
|---|---|
| Insignificant (0.00 to ±0.09) | 3 |
| Weak Significance (0.10 to ±0.29) | 7 |
| Moderate Significance (0.30 to ±0.49) | 16 |
| Strong Significance (0.50 to ±0.69) | 7 |
| Very Strong Significance (0.70 to ±0.89 | 2 |
| Perfect (0.10 to ±1.00) | 0 |

**Table 7: Aggregated test of significance for the studies.**

## 6. Discussion and Conclusion

This SLR aimed to understand the role of individual self-efficacy on IS security innovation adoption. The results highlighted that individual self-efficacy is a significant attribute of IS security innovation adoption. The SLR results of the distribution of studies by publication year suggest that researchers have started examining the effect of individual self-efficacy on IS security innovation adoption since 2004. This is the period where online social media and social networking became a mainstream concept with the launching of Facebook on February 2004. These social media emerge as a target for scams; exposing individual and organisational data at risk. More people put their personal information online, offering a huge opportunity for cyber criminals to exploit. Thus, IS security innovation adoption has speedily been under scrutiny since the rise of social media and researcher started examining individual self-efficacy as one of the key predictors for IS security innovation adoption.

Studies that examined the influence of individual self- efficacy for IS security innovation adoption has explored for different sample groups. The SLR findings showed that the research on the relationship between individual self-efficacy and IS security innovation adoption based their studies on convenience samples of both students and non-students. The findings indicate that approximately half of the reviewed literature used student subjects. Using student subjects for experimental research as a substitute for another group has been widely criticised for having little external validity and generalisability. The ethical concerns of student participation revolve mainly around the issue whether the participant serve with their own consent. Also, it has been argued that student samples are fundamentally biased in age, experience, and intellectual ability. However, the studies reviewed in the SLR provided no justification for their chosen subject sample nor did acknowledge any limitations for the use of a student sample. Hence, the effect of individual self- efficacy for IT security innovation adoption bears no significance for the difference in sample groups.

The results of SLR showed that the average sample size of the studies is approximately 350 participants. A study that has a sample size which is too small may have an unrealistic chance of yielding a useful information. Larger sample sizes have the obvious advantage of providing more data for researchers to work with and provide more accurate mean values and a smaller margin of error. Thus an appropriate determination of the sample size used in a study is a crucial step in the design of a study. The sample size used in the majority of the studies reviewed in the SLR deemed appropriate. This commends of the soundness of the selected studies for the SLR. In addition, it provides evidence on the correctness of the results of the reviewed studies that examine the relationship between individual self- efficacy and IT security innovation adoption.

In order to identify if culture moderates the relationship between individual self-efficacy and IS security innovation adoption, we explored the distribution of country of the reviewed studies in the SLR. Deans et al. [18] states that culture influences usage of IT in different countries. In a meta-analysis of TAM, Schepers and Wetzels [52] used western and non-western as a moderating factor in the context of culture. They divide the studies conducted in Europe, North America, Australia and New Zealand as western and the rest of world as non-western. The SLR represents a diverse culture which belongs to both western and non-western groups. Hence, the SLR indicates that the overall results of existing literature that considers the influence of individual self-efficacy for IS security innovation adoption is not biased towards one particular culture.

The SLR also explored the theoretical foundation exploited in examining individual self-efficacy for IS security innovation adoption by the reviewed studies. The result of the SLR identified PMT as the principal model. In a meta-analysis study, Floyd et al. [22] described PMT as one of the most powerful explanatory theories predicting individual intentions to adopt safeguard measures. PMT is useful in analysing and exploring recommended actions or behaviours to avert the consequences of threats such as IS security attacks. Apart from PMT, SLR identified SCT, TRA, and TPB as other models utilised in examining the effect of individual self-efficacy for IS security innovation adoption. SCT [7] posits that one's confidence in their ability to perform it a behaviour successfully will produce positive valued outcomes. The main tenet in the TRA is that an individual's behavioural intention in a specific context depends on attitude toward performing the target behaviour and on subjective norm. The TRA holds that the practical impact of subjective norm on the behavioural intention is that an individual may choose to perform a specific behaviour, even though it may not be favourable to him or her to do so [64]. TPB is an extension of TRA at the same time adopt the efficacy expectancies of SCT into consideration.

In this study, we identified if there is a difference in investigating individual self-efficacy for different types of security innovations. In order to analyse, we categorised IS security innovations as product and process to access the scenario. The results show that the average sample size used for IS security process innovation studies (373 participants) is higher than the product innovation studies (271 participants). One explanation is that process innovation involves replacing the entire system or work procedure, whereas product innovation does not involve change of an entire system. Also, it is evident from the results that most of IS security process innovation studies utilises individuals as subjects, whereas most of IS security product innovation studies employ students. One probable explanation could be that process innovation such IS security policies are mostly adopted in an organisational setting for which the sample subjects would most probably be non-students.

Finally, the SLR analysed the correlation co-efficient for the relationship between individual self-efficacy and IS security adoption behaviour to aggregate the tests of significance of the reviewed studies. In terms of the percentage, 92% of the studies found self-efficacy as significant (correlation value between ±0.10 to ±1.00) attribute in IS security innovation adoption. Also, approximately 71% of the studies we considered verified the association between self-efficacy and IS security adoption as moderate significance (correlation value between ±0.30 to ±0.49) or strong significance (correlation value between ±0.50 to ±0.69). Hedges and Olkin [31], Hameed and Counsell [24] and Hameed and Counsell [26] suggested that it would be within reason for a study to consider an established relationship to exist between two variables when a majority of prior studies had found statistically significant results. Hence, results of aggregated tests of significance indicate that individual self-efficacy is an important predictor of IS security innovation adoption.

This study offers several contributions to the IS security management literature. The study contributes to the field of IS security by empirically endorsing the influence of individual self-efficacy for IS security innovation adoption. Additionally, to recognise the current understanding of the subject, we gathered almost all their existing studies that examine individual's self- efficacy for IS security innovation adoption.

The most important theoretical implication is that this study using SLR verifies the significance of self- efficacy for IS security innovation adoption. Another key implication of this study is the importance of spreading IS security knowledge among the users for safeguarding IS assets. On one hand, knowledge has a simple positive effect on self-efficacy, which affects the individual's security behaviour. On the other hand, knowledge allows users to assess a security technology fairly and improve the quality of decision making. IS security literature has emphasised on the need to pay attention to security education, awareness and training initiatives and interventions. Therefore, we suggest that organizations create appropriate education, training and security awareness programs that ensure employees possesses up-to-date knowledge of IS security as well as facilitate conditions that will improve their individual self-efficacy regards IS threats. This study has certain limitations. The major limitation of this analysis was the inadequacy of studies that examined individual self- efficacy on IS security innovation adoption. The result of the SLR would be more accurate and better explained if analysed with more studies.

# APPENDIX

| SDY NAME | YER | SAM G | SAM S | CNTRY | Theories | INN TYP | COR |
|---|---|---|---|---|---|---|---|
| Herath and Rao (2009) | 2009 | ORG | 312 | USA | PMT, DT, DTPB | PRC | 0.51 |
| Ng et al. (2009) | 2009 | MIX | 134 | Singapore | HBM | PRD | 0.4 |
| Mohamed and Ahmad (2012) | 2012 | SDT | 340 | Malaysia | PMT, SCT | PRC | 0.419 |
| Son (2011) | 2011 | IND | 602 | USA | EMM, IMM | PRC | 0.23 |
| Workman et al. (2008) | 2008 | IND | 588 | USA | PMT | PRC | |
| Rhee et al. (2009) | 2009 | SDT | 415 | USA | SCT | PRC | 0.363 |
| Johnston and Warkentin (2010) | 2010 | MIX | 215 | USA | PMT | PRD | 0.342 |
| Bulgurcu et al. (2010) | 2010 | ORG | 464 | Canada | TPB, RCT | PRC | 0.395 |
| Yoon et al. (2012) | 2012 | SDT | 202 | South | PMT | PRC | 0.1 |
| Ifinedo (2012) | 2012 | IND | 124 | Canada | TPB, PMT | PRC | 0.32 |
| Ifinedo (2014) | 2014 | IND | 124 | Canada | TPB, SCT, SBT | PRC | 0.24 |
| Anderson and Agarwal (2010) | 2010 | IND | 594 | USA | PMT | PRD | 0.44 |
| Anderson and Agarwal (2010) | 2010 | IND | 101 | USA | PMT | PRD | 0.38 |
| Chou and Chien Chou (2016) | 2016 | IND | 505 | Taiwan | PMT | PRD | 0.05 |
| Warkentin et al. (2016) | 2016 | SDT | 253 | USA | PMT | PRD | 0.888 |
| Siponen et al. (2014) | 2014 | IND | 669 | Finland | TRA, CET | PRC | 0.243 |
| Tamjidyamcholo et al. (2013) | 2013 | IND | 138 | Malaysia | TRA, SCT | PRC | 0.566 |
| Lee et al. (2008) | 2008 | SDT | 273 | USA | PMT | PRD | 0.6 |
| Vance et al. (2012) | 2012 | IND | 210 | Finland | PMT | PRC | 0.47 |
| Chan et al. (2005) | 2005 | IND | 104 | Singapore | | PRC | 0.4 |
| Herath et al. (2014) | 2014 | SDT | 134 | USA | TAM, TTAT | PRD | -0.08 |
| Marett et al. (2011) | 2011 | SDT | 522 | USA | PMT | PRC | 0.51 |
| Lui and Hui (2011) | 2009 | SDT | 752 | China | TAM | PRD | 0.082 |
| Wei and Zhang (2008) | 2008 | SDT | 279 | China | TAM | PRC | 0.32 |
| Sun et al. (2016) | 2016 | SDT | 411 | Taiwan | | PRD | 0.52 |
| Sun et al. (2016) | 2016 | SDT | 411 | Taiwan | | PRD | 0.45 |
| Liang and Xue (2010) | 2010 | SDT | 152 | USA | TTAT | PRD | 0.283 |
| Dinev et al. (2009) | 2009 | SDT | 332 | USA | TPB | PRD | 0.39 |
| Dinev et al. (2009) | 2009 | SDT | 227 | South | TPB | PRD | 0.35 |
| Hanus and Wu (2016) | 2016 | SDT | 229 | USA | PMT | PRC | 0.65 |
| Lai et al. (2012) | 2012 | SDT | 117 | USA | CT | PRC | -0.186 |
| Meso et al. (2013) | 2013 | SDT | 77 | USA | PMT | PRD | 0.784 |
| Siponen et al. (2007) | 2007 | IND | 917 | Finland | PMT, DT, TRA | PRC | 0.407 |
| Tamjidyamcholo et al. (2013) | 2013 | SDT | 138 | Malaysia | PMT | PRC | 0.565 |
| Tsai et al. (2016) | 2016 | IND | 988 | USA | PMT | PRC | 0.26 |
| Chenoweth et al. (2009). | 2009 | IND | 204 | USA | PMT | PRD | |
| Crossler (2010) | 2010 | IND | 112 | USA | PMT | PRD | |
| D'Arcy and Hovav (2004) | 2004 | NON | | | DT | PRC | |
| Cox (2012) | 2012 | IND | 106 | USA | TPB | PRC | 0.43 |
| Lee et al. (2007) | 2007 | NON | | USA | PMT | PRD | |
| Milne et al. (2009) | 2009 | IND | 449 | USA | PMT, SCT | PRC | |
| Pahnila et al. (2007) | 2007 | IND | 917 | Finland | PMT, DT, TRA | PRC | |

[YER - Year], [SAM G - Sample Group: IND - Individual; ORG - Organisation; SDT - Student; MIX - Mixed; NON - None], [SAM S - Sample Size], [CNTRY - Country], [Theories: PMT - Protection Motivation Theory; TPB - Theory of Planned Behaviour; TRA - Theory of Reasoned Action; SCT - Social Cognitive Theory; DT - Deterrence Theory; TAM - Technology Acceptance Model; TTAT - Technology Threat Avoidance Theory; CET - Cognitive Evaluation Theory; CT - Coping Theory; DTPB - Decomposed Theory of Planned Behaviour; EMM - Extrinsic Motivational Model; HBM - Health Belief Model; IMM - Instrinsic Motivation Model; RCT - Rational Choice Theory; SBT - Social Bond Theory], [INN TYP - Innovation Type: PRC - Process; PRD - Product], [COR - Correlation]